\documentclass[12pt]{article}
\usepackage{graphics}
\usepackage{latexsym}
\usepackage{pgf}
\newcommand{\beq}{\begin{equation}}
\newcommand{\eeq}{\end{equation}}

\newcommand{\beqs}{\begin{eqnarray}}
\newcommand{\eeqs}{\end{eqnarray}}

\begin{document}
\bibliographystyle{h-physrev}
\begin{titlepage}
\vskip 2.5cm
\begin{center}
{\LARGE Aspects of Integrability in $\mathcal{N}$ =4 SYM}\\
\vskip 1.2cm
{\Large Abhishek Agarwal} \
\vskip 1.3cm {\Large Physics Department}\\
{\large City College of the CUNY}\\
{\large New York, NY 10031}\\
abhishek@sci.ccny.cuny.edu \vskip 0.7cm
\end{center}
\vspace{3.14cm}
\begin{abstract}
Various recently developed connections between supersymmetric
Yang-Mills theories in four dimensions and  two dimensional
integrable systems serve as crucial ingredients in improving our
understanding of the AdS/CFT correspondence. In this review, we
highlight some connections between superconformal four dimensional
Yang-Mills theory and various integrable systems. In particular, we
focus on the role of Yangian symmetries in studying the gauge theory
dual of closed string excitations. We also briefly review how the
gauge theory connects to Calogero models and open quantum spin
chains through the study of the gauge theory duals of D3 branes and
open strings ending on them. This invited review, written for Modern
Physics Letters-A,  is based on a seminar given at the Institute of
Advanced Study, Princeton.
\end{abstract}

\end{titlepage}
\section{Introduction and Summary}
The proposed duality between Yang-Mills theories and string theories
has been a fascinating avenue for explorations since Maldacena's '97
conjecture\cite{malda}. The duality relating the weak-coupling
regime of  a ten dimensional string theory to the strongly coupled
sector of a four dimensional super Yang-Mills (SYM) theory has been
notoriously hard to test. However, some fascinating new  discoveries
have led to a remarkable amount of progress on the problem. The
basic insight that has allowed on to make some headway into the
problem is that both the ten dimensional string theory and the four
dimensional  gauge theory appear to be two dimensional integrable
models in disguise! That there is an underlying two dimensional
dynamical system underlying the string theory is perhaps no so
surprising, as one is naturally led to a world sheet sigma model in
the analysis of the string dynamics. What is rather counterintuitive
is that for the purposes of computing it's spectrum, the four
dimensional gauge theory can be regarded as a two dimensional
integrable system as well. The closest thing that one has to a
spectrum of masses for this four dimensional conformal field theory
is the spectrum of anomalous dimensions i.e the spectrum of the
dilatation operator.  In a remarkable paper, Minahan and Zarembo
\cite{minz} were able to show that the large $N$ dilatation operator
of the gauge theory, when restricted to the sector of operators
built out of scalars can be regarded as the Hamiltonian of an
integrable quantum spin chain. Subsequent analyses have shown that
this is a feature that is  restricted neither to the one loop level
nor to operators built out of scalars\cite{bs1,bs2,bs3}.

Since, the original results of Minahan and Zarembo the understanding
of integrability in the gauge theory has only deepened. Due to
extremely impressive gauge theory computations\cite{bern1}, the spin
chain describing the dilatation operator of the Yang-Mills theory is
now known up to four loop orders. To the extent that we have a
formula for the spin chain, there is a great deal of evidence that
it is integrable\cite{bs4}.

Given a quantum spin chain, the surest test of its integrability can
probably be formulated in terms of algebraically independent
commuting higher charges and/or the existence of underlying quantum group
symmetries which can ultimately be understood in terms of an
appropriate transfer matrix. Such a detailed microscopic
understanding of the integrability of the quantum spin chains
emerging from the Yang-Mills theory exist at the first few orders in
perturbation theory. This has to do with the fact the we know the
Hamiltonian of the quantum spin chain relevant to $\mathcal{N}$ =4
SYM only to the first few orders in perturbation theory. However, a
lot of progress is possible even without the explicit knowledge of
the Hamiltonian. Beisert has recently shown that the fact that the
underlying gauge theory is maximally supersymmetric, i.e possess a
$psu(2,2|4)$ invariance, can be used to almost uniquely fix the two
particle scattering matrix of its dilatation operator, even without
the knowledge of the explicit form of the appropriate spin chain
Hamiltonian. It turned out that by considering the superconformal
group as a particular contraction of one of its non-central
extensions, the dispersion relation of the underlying spin chain
could be uniquely fixed. Furthermore, the two particle $S$ matrix is
determined up to a single undetermined function/phase, when one uses
this approach\cite{bss1,bss2}. He was also able to show that this
particular two body $S$ matrix also satisfies the Yang-Baxter
algebra: a necessary condition for integrability. This result brings
one very close to an exact solution of the large $N$ limit of the
gauge theory, as far as its spectrum is concerned. Indeed, Beisert's
result on the $S$ matrix and the undetermined phase, have been
largely constrained now and some extrapolations of the results to
the strong coupling regime have also been successfully carried
out\cite{bs4,kleb1}. These exciting chain of developments allow one
to make some very concrete statements about the validity of the
AdS/CFT correspondence.

However, various open questions still remain. The fact that the $S$
matrix for the spin chain underlying the gauge theory satisfies the
factorization conditions, implies that the underlying Hamiltonian
has a very good chance of being integrable. If that is indeed the
case, then it should be, at least in principle, possible to
understand its integrability in terms of symmetries. While that is
not possible yet in complete generality, important steps in that
direction have already been taken. In the first part of the review,
we shall briefly sketch out the nature of some of the non-trivial
conservation laws and symmetry principles that can help us
understand the perturbative integrability of the gauge theory.

Much of the success regarding the understanding of integrability of
$\mathcal{N}$=4 SYM, alluded to above, has had to do with studies of
gauge theory operators \beq \mathcal{O} =
\mbox{Tr}(\mathcal{W}_1\cdots \mathcal{W}_m).\eeq i.e. operators
constructed out of traces. $\mathcal{W}_i$s stand for generic
Yang-Mills fields. However, the gauge theory also possess many other
degrees of freedom, such as operators built out of multiple traces,
determinants and sub-determinants etc. Such degrees of freedom allow
us to probe various corners of the AdS/CFT correspondence, which are
not necessarily accessible by straightforward applications of the
techniques developed in the studies of single trace operators. Gauge
theory operators which are dual to D branes, for instance, cannot be
built from traces. The second part of the review will be devoted to
a brief overview of some of the developments in our understanding of
Baryonic/D-Brane degrees of freedom on the gauge theory side and the
relevant integrable systems.

This brief overview is by no means meant to be a complete synopsis
of the rapidly growing body of exiting work relating supersymmetric
Yang-Mills theories, two dimensional integrable systems and string
theory. For instance, most of the recent developments regarding the
 understanding of the  world-sheet $S$ matrices of the gauge and string
theories will not be covered in the article. However, given our
knowledge of world-sheet scattering in the gauge theory, the
spectrum of single trace operators to all orders in perturbation
theory, etc, it might be argued that the time is ripe for a more
detailed understanding of integrability in the gauge theory. The
review focusses on some particular aspects of integrability in
$\mathcal{N}$ =4 SYM  about which our understanding remains
incomplete. In particular we will focus on the role of Yangian
symmetries in the gauge theory and the gauge theoretic description
of D-Brane degrees of freedom. The review is based on a seminar
given at the Institute for Advanced Studies, Princeton.

\section{Gauge Theory Set Up:}
$\mathcal{N}$ =4 supersymmetric Yang-Mills theory in four dimensions is a conformal field theory with a
 vanishing beta function. The closest thing  to the spectrum of masses for this truly massless theory is
 its spectrum of anomalous dimensions.  For the purposes of computing the
 spectrum it is extremely convenient to regard the local composite
 operators, formed out of considering fundamental constituent
 Yang-Mills degrees of freedom at the same spacetime point inside a
 color trace, as the states of an appropriate quantum spin chain.
 For instance, if we were to restrict ourselves to the closed $su(2)$
sub-sector of the gauge theory made up of two complex scalars $Z,W$,
which are charged under different $U(1)$ subgroups of the $so(6)$ R
symmetry group, one gets $su(2)$ spin chains with spins in the
fundamental representation. \beq Tr(ZZWWZ) \Leftrightarrow
(\Uparrow\Uparrow\Downarrow\Downarrow\Uparrow).\eeq

 As one turns on the 't Hooft coupling of the gauge theory, $\lambda
$, the effects of various interaction vertices, will lead to the
this operator mixing with various other operators that carry the
same quantum numbers. However, in the large $N$ limit, since,
spitting of single traces into multiple traces is suppressed, all
that the gauge theory vertex insertions can bring about is a
rearrangement of the fields inside the trace. One can thus think of
the effect of the vertex insertion i.e the action of the dilatation
operator $D$,  as the action of an appropriate spin chain
Hamiltonian on the state of interest. In the $su(2)$ sector of the
gauge theory, the dilatation operator takes on the following form up
to the three loop level\cite{ser1,ser2}. \beqs D =
\sum_i((1-P_{i.i+1}) + \lambda
((1-P_{i.i+2})-4(1-P_{i.i+1}))\hspace{2cm}\nonumber\\
+ \lambda ^2(28(I-P_{i,i+1}) -8(I-P_{i,i+2})
-2(P_{i,i+3}P_{i+1,i+2}-P_{i,i+2}P_{i+1,i+3}))+
\cdots)\label{dilop}\eeqs $P_{i,j}$ is the permutation operator that
exchanges the spins at the sites $i$ and $j$. The  one loop $su(2)$
dilatation operator (the leading therm in the above formula), is of
course nothing but the celebrated ferromagnetic Heisenberg spin
chain, a well known integrable many-body system, providing us with a
first contact between integrable structures and the gauge theory.

Interpreting the gauge theory dilatation operator as a quantum spin
chain greatly facilitates understanding many aspects of the AdS/CFT
correspondence. Most importantly,  it is possible to apply Bethe
ansatz techniques to compute the gauge theory spectrum by purely
algebraic means. As mentioned earlier, much progress has been
achieved understanding the higher loop spectrum of the gauge theory
in various exciting recent papers, which we shall not review in the
present work. We shall instead now focus on the present
understanding of the symmetries and conservation laws responsible
for integrability on the gauge theory side.

\section{Symmetries and Conservation Laws:} To motivate the
non-trivial conservation laws leading to our present understanding
of the perturbative integrability of the gauge theory, it is
instructive to look at the one loop $su(2)$ dilatation operator, i.e
the Heisenberg Hamiltonian, which we re-write as, \beq D_1 = \sum_i
(1- S^a_b(i)S^b_a(i+1)).\eeq $S^a_b(i)$ is the Weyl operator that
acts as $|a><b|$ at the lattice site $i$. The Hamiltonian has an
obvious $su(2)$ symmetry, generated by \beq (\mathcal{Q}^0)^a_b =
\sum_i S^a_b(i).\eeq However, this does not, by itself, generate
enough conservation laws to render the system solvable. However
there exists another non-local charge that does commute with the
Hamiltonian. \beq (\mathcal{Q}^1)^a_b =
\frac{1}{2}\sum_{i<j}\theta(i,j)\left(S^a_k(i)S^k_b(j)-
(i\leftrightarrow j)\right).\eeq For the Heisenberg model, $\theta
(i,j) = 1 \forall i,j$, however for more complicated spin chains
that we shall consider, $\theta $ can have a more complicated
functional dependence. These two charges, known as the Yangian
charges, are ultimately responsible for the integrability of the
system. It turns out, that by forming repeated commutators, of the
charges with each other, one can generate an infinite tower of
algebraically independent charges, all of which commute with the
Heisenberg Hamiltonian. The symmetry underlying the dynamical system
is not a Lie algebraic one, as the charges do not close on each
other under commutation. As a matter of fact they form a Hopf
algebra with co-products given by
\begin{eqnarray}\Delta(\mathcal{Q}^0)^a_b  = (\mathcal{Q}^0)^a_b
\otimes \mathcal{I} + \mathcal{I}\otimes (\mathcal{Q}^0)^a_b\\
\Delta(\mathcal{Q}^1)^a_b  = (\mathcal{Q}^1)^a_b \otimes
\mathcal{I}+ \mathcal{I}\otimes  (\mathcal{Q}^0)^a_b
\nonumber \\
+( (\mathcal{Q}^0)^a_d\otimes  (\mathcal{Q}^0)^d_b
-(\mathcal{Q}^0)^d_b\otimes (\mathcal{Q}^0)^a_d)\nonumber
\end{eqnarray}
The co-product must be an algebra homomorphism. Physically, this may
be understood as the requirement that the algebra not change, if one
adds more spins to the spin chain state. The requirement restricts
the function $\theta $ which is required to satisfy the following
constraint, also known as the Serre relation \beq
\theta(j,k)\theta(j,n) + \theta(j,k)\theta(n,k) -
\theta(j,n)\theta(n,k) = \theta(j,k),\eeq which is trivially
satisfied if $\theta  =1$

As far as the Heisenberg model is concerned, all the algebraically
independent charges obtained from the iterated commutators of the
two Yangian charges can be captured in a ($2\times 2$) transfer
matrix, which can be expressed as a path ordered exponential of the
Weyl matrix. \beq T^a_b(u) =
\left[e^{\wp\frac{1}{u}\sum_iS(i)}\right]^a_b = \sum_i
\frac{1}{u^n}(T_n)^a_b\eeq $u$ is the spectral parameter, and the
model expansion around $u\rightarrow \infty $ reads as  \beq
T(u)^a_b = \mathcal{I} + \frac{1}{u}(\mathcal{Q}^0)^a_b +
\frac{1}{u^2}(\mathcal{Q}^1)^a_b + \cdots\eeq The commutation
relations between the various modes of the transfer matrix give us
the famous Yang-Baxter algebra \beq [T^{ab}_s, T^{cd}_{p+1}] -
[T^{ab}_{p+1}, T^{cd}_{s}] =\left( T^{cb}_{p} T^{ad}_{s} -
T^{cb}_{s} T^{ad}_{p}\right)\eeq  while the co-product is reflected
as \beq \Delta T^{ab} = \sum_d T^{ad}T^{db}\eeq The Heisenberg
Hamiltonian is an element of the center of the Yang-Baxter algebra
and it commutes with $T$ i.e  \beq[D_1, T^{ab}_n]=0.\eeq
\footnote{The above statement is strictly true if the spin chains
are infinitely long. For chains of finite length, one has to
consider an appropriate quotient of the algebra of the spin-chain observables by
a proper ideal to realize the non-local conservation laws. See for
example \cite{aa1}}Perhaps a more familiar way to write the
Yang-Baxter algebra is by using the $R$ matrix \beq
R(u-v)(T(u)\otimes \mathcal{I})(\mathcal{I}\otimes T(v))=
(\mathcal{I}\otimes T(v))(T(u)\otimes \mathcal{I})R(u-v) \eeq where,
\beq R = u\mathcal{I} - P.\eeq The key point here is that the
transfer matrix and the Yangian charges carry precisely the same
information. As a matter of fact, the Yang-Baxter relations for the
matrix elements of the transfer matrix are nothing but the Serre
relations in disguise. A detailed algebraic construction relating
all the matrix elements of transfer matrix of the Heisenberg model
to the Yangian charges, along with a great deal of pedagogical
exposition, may be found for example in\cite{ber,ge}. Given a
transfer matrix satisfying the Yang-Baxter relations, a purely
algebraic way of diagonalizing the entire transfer matrix also
exists due to the work of Faddeev and collaborators \cite{fadrev}.
Thus, to sum up the lessons from the Heisenberg model, if one is
given a quantum spin chain, a fairly conclusive way to establish its
integrability consists of showing the existence of Yangian charges
that commute with it.

That there might be an underlying Yangian symmetry in the gauge
thoery was first suggested in \cite{witten}.  We can now focus on
how the above understanding of Yangian symmetries applies to the
gauge theory. If one were to focus on the $su(2)$ sector, then one
sees that, there are corrections to the Heisenberg Hamiltonian, with
the range of the spin chain increasing with the loop order. The
three loop Hamiltonian, for instance is given by ({\ref{dilop}).
Given the knowledge of the Dilatation generator up to a given loop
order, (say $\mathcal{O}(\lambda ^m)$), $D_m$, one would like to
find $\lambda $ dependent deformations of the Yangian charges, that
commute with the Hamiltonian up to terms of $\mathcal{O}(\lambda
^{m+1})$. In the $su(2)$ sector, we are aided by the realization
that the fist Yangian charge, i.e the $su(2)$ generator remains an
exact symmetry of the dilatation operator to any order, since the
$su(2)$ 'flavor' symmetry is manifestly realized in perturbation
theory. To find the second charge, we can use the fact that the
Serre relations have non-trivial solutions. For example, \beq \theta
(i,j) = \frac{t^{-i}}{t^{-i} - t^{-j}} \eeq solves the Serre
relations, for arbitrary, complex values of 't'. In particular
choosing \beq t = \sum_{n=1}^\infty  c_n \lambda ^n, c_1=1, c_2=-3,
c_3=14 \eeq establishes the three loop Yangian invariance in the
$su(2)$ sector of the gauge theory. In explicit form, the second
Yangian charge looks like
\begin{eqnarray}
(\mathcal{Q}^1)^a_b &=& \sum_{i<j}S^a_d(i)S^d_b(j)
+\lambda \sum_i S^a_d(i)S^d_b(i+1)\nonumber \\
&  &-\lambda ^2 \sum_i\left(2S^a_d(i)S^d_b(i+1) -
S^a_d(i)S^d_b(i+2)\right)+\lambda ^3\sum_i\left(9S^a_d(i)S^d_b(i+1)
+
S^a_d(i)S^d_b(i+3)\right)\nonumber \\
& &-((a,d)\leftrightarrow (d,b))\nonumber \end{eqnarray} This
construction establishes three loop integrability in the $su(2)$
sector of the gauge theory. These charges were computed order by
order in perturbation theory in \cite{aa1} For this particular
sector, one can also appeal to other constructions. For instance, it
was shown in\cite{ser1} that the three loop $su(2)$ dilatation
operator of the gauge theory can be embedded in the Inozemtsev spin
chain, which does have a Yangian symmetry. The Yangian charges,
reported above, can also be derived, by considering the appropriate
limits of the Yangian charges of the Inozemtsev model.

In proceeding beyond the $su(2)$ sector, one has to take into
account that even the flavor symmetry, (usually generated by the
first Yangian charge), is not always manifest. Nevertheless, in such
sectors of the gauge theory, constructions similar to the above
discussion are indeed possible, at least to the first two orders in
perturbation theory. For instance, for the $su(1|1)$ and $su(2|1)$
sectors, some recent results have been possible in \cite{aa2,bz}.
Finally, it is worth mentioning, that, as far the $S$ matrix of the
gauge spin chain is concerned, it is basically known to all loops in
perturbation theory, up to a single overall phase. This $S$ matrix
has recently been shown to possess Yangian invariance by Beisert
\cite{byan}. For related work, see also\cite{toryan}. One would of
course expect that the Yangian symmetry of the $S$ matrix would
intimately be related to that of the Hamiltonian discussed above. It
remains a fascinating subject of research to explain and tie
together all these glimpses of an underlying Yangian structure
obtained in the gauge theory so far.
\section{Protected Operators and Calogero Models:}
A different class of integrable dynamical systems emerge as the
effective Hamlitonians of the gauge theory when one focusses on
special operators that are protected against renormalization. One of
the best studied example is that of the dynamics of $\frac{1}{2}$
BPS, operators of the gauge theory, which are scalars $Z$ charged
under a $U(1)$ subgroup of the $R$ symmetry group. If one considered
the theory on $R\times S^3$, then the non-renormalization condition
reduces the dynamics of the this sector of the theory to that of the
zero models of these scalars. \beq Tr(Z)^J \Leftrightarrow
Tr(A^{\dagger})^J|0>, H_{SYM} \Leftrightarrow Tr(A^\dagger A)\eeq In
other words, the operators, distinguished by theory total $R$ charge
can be mapped to states of Hamiltonian matrix model, and the gauge
theory effective Hamiltonian takes on the form of a gauged matrix
Harmonic oscillator. The Hamiltonian, has a residual $U(N)$ gauge
invariance. One can remove that by going to the space of the
eigenvalues of the matrix model. As is well known from the study of
matrix models, this turns the system into a system of $N$
non-relativistic Fermions, which are otherwise free, and described
by the Hamiltonian \beq H = \sum_i\left(-\frac{\partial ^2}{\partial
x_i^2} + x_i^2\right)\eeq Recently, in a remarkable paper\cite{llm},
it was shown that the phase space of the free Fermion system, is
precisely the same as the space of all supergravity geometries that
are also half BPS, preserve an $O(4)\times O(4)$ isometry and are
asymptotically AdS.

Inspired by this result an attempt was made in\cite{aaa} to
understand the gauge fixed dynamics of a more general class of
operators within the gauge theory, which are protected, but not
necessarily because of supersymmetry reasons. The motivation being
that these will then be the prototypical operators that have a
chance of existing in theories without as much supersymmetry as
$\mathcal{N}$ =4 SYM.

One can focus on gauge theory operators built out of two Yang-Mills
fields, which we take to be a complex scalar $X$ and a Fermion $\Psi
^1$, which together make up the closed $su(1|1)$ subsector of the
gauge theory. Restricting to operators that are protected implies
that the effective Hamiltonian will be a sum of two Harmonic
oscillators \beq H = Tr(A^\dagger A + B^{\dagger}  B), A,B
\leftrightarrow Z,\Psi^{1} \eeq To gauge fix it, we can go to a
basis such that, \beq A = \frac{1}{2}(X + iP) , X = U^\dagger x U, x
= diag(x_1 \cdots x_N)\eeq and we can denote the impurity of
Fermionic fields in this basis by \beq (b )^i_j = (UB U^\dagger
)^i_j \eeq The Hamiltonian in this basis takes on the form \beq H =
\sum_i\left(-\frac{\partial }{\partial x_i^2} + x_i^2\right) +
\sum_{i\neq j}\left( \frac{\mathcal{L}^i_j\mathcal{L}^j_i}{(x_i -
x_j)^2}\right) + tr b^{\dagger}b \eeq where, \beq \mathcal{L}^i_j =
U^i_m\frac{\partial }{\partial U^m_j}, [\mathcal{L}^i_j,
\mathcal{L}^k_l] = \delta ^k_j \mathcal{L}^i_l -\delta ^i_l
\mathcal{L}^k_j.\eeq $\mathcal{L}^i_j$ are generators of the $U(N)$
rotations. Because, we have a system of two matrices, gauge fixing,
does not completely remove the $U(N)$ degree of freedom,
nevertheless i writing the Hamiltonian in the above form, we have
been able to separate out the 'radial' $x_i$ and the angular
$\mathcal{L}^i_j$ degrees of freedom. The Hamiltonian describes a
system of particles on the line, that carry an  internal spin $U(N)$
degree of freedom, and interact with each other by exchanging spins
on top of having an inverse square interaction. In other words it is
nothing but the  $SU(N)$ generalization of the celebrated Calogero
model. The states are functions of the kind \beq
 \Psi^{i_1 \cdots i_n}_{j_i \cdots j_n}(x)\Pi _{k= 1}^n (b^{\dagger
}) ^{j_k}_{i_k} |0> \eeq , where, $|0>$ is the vacuum with no
Fermions. It is important to note that because of the global $U(N)$
invariance of the problem, the $U(N)$ generators may be expressed in
terms of the $b$ fields as \beq \mathcal{L}^i_j =
 \sum _\beta\left( (b^{\dagger})^i_l(b^{ })^l_j -
(b^{\dagger  })^l_j(b^{})^i_l\right) .\eeq Not all states of the
Matrix model are protected gauge theory operators, however one could
focus on states such as
\beq\frac{1}{\sqrt{N^{n_1+1}}}tr\left((A^{\dagger })^{n_1}B^{\dagger
}\right) \frac{1}{\sqrt{N^{n_2}}}tr \left((A^\dagger )^{n_2}\right)
\cdots  \frac{1}{\sqrt{N^{n_i+1}}}tr\left((A^{\dagger
})^{n_i}B^{\dagger }\right)|0>\eeq These states have the curious
property, that they are  protected in the Large $N$ limit, while for
finite $N$ they correspond to  like BMN Like Near Chiral Primariry
operators. Thus in the large $N$ limit, they are protected, without
being protected due to supersymmetry.  In the gauge fixed language
they take on the form \beq \prod_m\Psi(x_1 \cdots x_N)^{i_1 \cdots
i_m}(b^{\dagger})_{i_1}\cdots (b^{\dagger })_{i_m}|0> +
O(\frac{1}{N}), (b^{\dagger \alpha })_i = (b^{\dagger \alpha
})^i_i.\eeq The spin degrees of freedom, Of which there were
approximately $N$, in the Calogero model, now reduce to just two.
These correspond to whether or not one does or does not have a
fermionic excitation within a trace. A detailed derivation of how
this comes about is given in\cite{aaa}, however, the key result is
that on these states of interest, the spin-spin interaction term can
be expressed as \beq  \mathcal{L}^i_j\mathcal{L}^j_i= \frac{1}{2}(1
- \Pi_{i,j}),\eeq where, $\Pi$ is a (graded) permutation operator
for the spin degrees of freedom.  The Hamiltonian, in turn, can be
written as \beq  D = \sum_i\left(-\frac{1}{2}\frac{\partial
}{\partial x_i^2} + b^{\dagger i}b_i + \frac{1}{2}x_i^2\right) +
\frac{1}{2}\sum_{i\neq j}\left( \frac{1-\Pi_{i,j}}{(x_i -
x_j)^2}\right).\label{cal}\eeq i.e the Hamiltonian of the  rational
Super-Calogero Model.

The Free fermion system corresponding to half BPS operators is a
special case of the Calogero system, it simply corresponds to states
with no impurity, $'b'$, type excitations. A detailed description of the
spectrum and  degeneracies of the Calogero model in terms of Young
diagrams has also been carried out in\cite{aaa}.

Since, the Calogero model, which is an integrable system, gives us a
window into the strong coupling dynamics of the gauge theory, it is
instructive to look at the algebraic structure underlying its
integrability and compare it with the ones known to be present in
perturbative, weakly coupled, SYM analysis.

The Lax operator for the Calogero system given in({\ref{cal}) can be
expressed as\cite{ak}
 \beq L_{j,k} = \delta _{j,k}\frac{\partial }{\partial x_j} + \hbar
(1-\delta _{j,k}) \theta(j,k)\Pi_{j,k}\eeq and \beq \theta(j,k) =
\frac{e^{-\frac{\hbar}{2}(x_i - x_j)}}{\sinh \frac{\hbar}{2}(x_i -
x_j)}\eeq This form of the Lax operator is more general than what is
needed, as a matter of fact, only the $\hbar \rightarrow 0$ limit
corresponds to the rational Calogero model given above. However, it
is instructive to keep this general form, with the fictitious
parameter $\hbar $ arbitrary. The non-local conserved charges for
the model can be constructed as \beq T_{n}^{ab} = \sum _{j,k}
S^{ab}(j)(L^n)_{j,k}\eeq and explicit computations show that
$T^{ii}_n$ are conserved. The Hamiltonian that commutes with the
non-local charges is expressible as \beq H_\hbar =
\frac{1}{2}\sum_{j,k}(-\partial ^2_j + x_j^2 + b^\dagger (j)b(j) +
\hbar \Pi _{j,k}\partial _j \theta(j,k) + \hbar
^2\theta_{j,k}\theta_{k,j}).\eeq The algebra of charges can be
computed to be \begin{eqnarray}[T^{ab}_s, T^{cd}_{p+1}]_\pm -
[T^{ab}_{p+1}, T^{cd}_{s}]_\pm = \hbar (-1)^{\epsilon (c) \epsilon
(a) + \epsilon (c) \epsilon (b) + \epsilon (b) \epsilon (a)}\left(
T^{cb}_{p} T^{ad}_{s} - T^{cb}_{s}
T^{ad}_{p}\right)\nonumber\end{eqnarray} $\epsilon (0) = 0,
\epsilon(1) = 1$. This is nothing but the supersymmetric Yangian
algeba encountered previously in the discussion of the spin chains.
It is important to note that the non-linearity of the Yang-Baxter
algebra is proportional to $\hbar$, and that in the $\hbar
\rightarrow 0$ limit, which is what is relevant for the Yang-Mills
theory, the Yangian algera degenerates into a loop algebra, \beq
[T^{ab}_s, T^{cd}_{p}]_\pm = \delta_{b,c} T^{ad}_{p+s} -
(-1)^{(\epsilon (a) + \epsilon (b))(\epsilon (c) + \epsilon
(d)}\delta_{a,d} T^{cb}_{p+s}.\eeq It has been known for sometime,
that loop algebras can be regarded as contractions or classical
limits of Yangian algebras\cite{ber}. Given the appearance of the
Yangian symmetry in perturbative gauge theory analyses, and that of
its classical limit in the small window into the strong coupling
sector provided by the Calogero model, it seems suggestive that a
contraction of the symmetry algebra takes place as one moves frmo
the weak to the strongly coupled regime of the gauge theory. It
would be extremely exciting if this possibility be explored further.

It may also be noted that for the rational super Calogero model, the
supercharges, and all the mutually commuting Hamiltonians can be
embedded in the loop algebra. \beq T^{21}_1 = Q, T^{12}_1 =
Q^\dagger,  H = [ T^{21}_1, T^{12}_1]_+\eeq. The higher conserved
Hamiltonians are given as \beq H_{n+m} = [ T^{12}_n, T^{21}_m]_+\eeq
while \beq [T^{11}_m, T^{11}_n] = [T^{11}_m, T^{22}_n] = [T^{22}_m,
T^{22}_n] = 0 \forall m,n\eeq The discussion here is only a brief
summary of some of the  connections between in the Yang-Mills theory
and Calogero systems presented in\cite{aaa}. For example it is
possible to extend the connection to the closed $su(2|3)$ sector of
the gauge theory which involves more than just two SYM fields. We
shall refer to\cite{aaa} for further reading.
\section{Open Spin Chains and Branes in SYM}
Apart from operators formed out of traces and products of traces of
Yang-Mills fields the gauge theory also has Baryonic operators whose
bare engineering dimensions are of $O(N)$. The simplest such
operator is \beq \mathcal{O} = \epsilon _{i_1 \cdots i_N}\epsilon
^{j_1 \cdots j_N}(Z^{i_1}_{j_1} \cdots Z^{i_N}_{J_N}).\eeq This
operator, being made up of $Z$ fields only is half BPS and
protected. From the point of view of the AdS/CFT correspondence,
such operators are important to study as they provide us with the
gauge theory analogs of D-Brane excitations. The above operator, for
example, is the SYM dual of a D3 brane. One can also construct the
dual of an open string ending ending on the giant graviton, simply
by removing one of the $Z$ fields from within the trace, and
replacing it by a matrix product of local Yang-Mills fields. \beq
\mathcal{O} \rightarrow \epsilon _{i_1 \cdots i_{N-1}
i_N}\epsilon^{j_1 \cdots j_{N-1}j_N}((Z^{i_1}_{j_1} \cdots
Z^{i_{N-1}}_{J_{N-1}})(WZZWZWW\cdots W)^{i_N}_{j_N}.\eeq One may ask
whether or not it makes sense to study the dynamics of these non-BPS
excitations as those of open quantum spin chains, with the string of
of Yang-Mills fields that replaces the $Z$ field playing the role of
an open quantum spin chain. These being operators of $O(N)$, and
keeping in mind that in the study of closed chains the spin chain
only emerged in the large $N$ limit, one has to carefully
investigate whether a sensible large $N$ limit leading to a spin
chain description of these operators is possible. That it is indeed
possible at the one loop level was shown by Berenstein and
collaborators in \cite{dber1}. In the $su(2)$ sector, the relevant
spin chain is given by \beq
 D_1 = \sum_{l=1}^{L-1}(\lambda )(I - P_{l,l+1}) + q_1^Z + q_1^L.\eeq

In the above formula, we have identifies all the $L$ fields lying
between the two boundary $W$ fields to be the spin chain. The 'bulk'
Hamiltonian of the spin chain is the same as the one for closed spin
chains, but the interactions between the spins in the chain with the
ones in the determinant introduce some boundary terms, denoted by
the $q$'s.$q^Z_i$ is a projection operator, that checks is the spin
at the $i$Ith site is equal to $Z$. If it is not, then it
annihilates the site, otherwise, it just acts as the identity. Put
differently, the present of the brane imposes Dirichlet boundary
conditions which dictate that the first and the last fields in the
chain cannot be $Z$.

This  spin chain is integrable, and it can be solved by Bethe ansatz
techniques. Its ground state is given by an the state \beq |0> =
\epsilon_{i_i \cdots i_{N-1}i_N}\epsilon^{j_i \cdots
j_{N-1}j_N}(Z^{i_1}_{j_1}\cdots Z^{i_{N-1}}_{j_{N-1}})(WWWW\cdots
WWW)^{i_N}_{j_{N}}.\eeq An eigenstate with two magnons can be
constructed as \beq|\Psi_2> = \sum_{x<y}\Psi(x,y)|x,y>\eeq with \beq
\Psi (x,y) = \sum _p \sigma (p)A(k_i,k_2)e^{i(k_1x_1 +k_2x_2)}.\eeq

In the above equation $x,y$ denote the positions of the flipped
spins i.e the $Z$ fields. The sum extends over all negations of the
momenta, and $\sigma $ in the sum indicates that we add a negative
sign each time the momenta are negated or permuted. i.e we make a
superposition of the incoming and outgoing plane waves. Unlike the
case of closed spin chains one has to account for the effect of
scattering from the boundary in addition to the mutual scattering of
the magnons. The Bethe equations, derived by following the magnons
around the spin chain and requiring that nothing change in doing so,
determine the momenta by the equations\beq \frac{\alpha (k_i)\beta
(k_i)}{\alpha (-k_i)\beta (-k_i)} = \prod_{j\neq
i}\frac{S(-k_i,k_j)}{S(k_i,k_j)},\eeq where $\alpha$ and $\beta $
encode the information about scattering from the boundary and $S$ is
the two magnon 'bulk' scattering matrix, given by:
\begin{eqnarray} \alpha (-k) = 1,
\beta (k) &=& e^{i(L+1)k},\end{eqnarray}  \beq S(k_1,k_2) =
1-2e^{ik_2} + e^{i(k_1+k_2)}.\eeq The energy, in terms of the
momenta is obtained by the dispersion relation \beq E = 4\lambda
\sum_i \left(\sin ^2(\frac{k_i}{2})\right).\eeq

One might ask if the situation changes at higher loops. It turns out
that the spin chain description of the non-BPS excitations around
the giant graviton background continues to hold at higher loops as
well. Furthermore, the 'bulk' Hamiltonian of the two loop spin chain
is precisely what one had in the closed string sector. The crucial
question has to do with the nature of the boundary interactions. In
the one loop case, the boundary Hamiltonian $q_1, q_L$ came about
from the condition that the boundary fields cannot be $Z$ i.e. the
Dirichlet boundary conditions. At two loops the situation is more
subtle and non-trivial boundary terms stemming from the interaction
of the bulk and boundary degrees of freedom. The two loop spin chain
Hamiltonian and the appropriate boundary terms have recently been
computed in \cite{maldaopen} and the result is:
 \begin{eqnarray}D_2
=\sum_{l=1}^{L-1}(1-4\lambda)(I-P_{l,l+1}) +
\sum_{l=1}^{L-2}\lambda(I-P_{l,l+2})\nonumber \\
+(1-2\lambda)q_1^z + \lambda q_2^Z + (1-2\lambda)q_L^Z + \lambda
q_{L-1}^Z.\label{open22}
\end{eqnarray}
As a historical aside, it is worth noting that in\cite{aaopen},
where a first attempt was made at computing the two loop open-chain
Hamiltonian, the result reported is the same as the one obtained if
one assumes that  Dirichlet boundary conditions are the only
source of boundary interactions at two loops. The bulk Hamiltonian
 in \cite{aaopen} is the same as (\ref{open22}), however the
boundary Hamiltonian is $(1-4\lambda)q_1^z + \lambda q_2^Z +
(1-4\lambda)q_L^Z + \lambda q_{L-1}^Z,$ which differs from the one
above by a factor of two in one of the terms. This boundary
Hamiltonian does not account for  a subtle but extremely crucial
contribution of a particular class of Feynman diagrams, which,
though apparently sub-leading order in $\frac{1}{N}$, end up giving
$\mathcal{O}(1)$ contributions due to contributions from the overall
scaling dimensions of the operators involved\cite{maldaopen}. A
careful analysis of the Bethe ansatz applied to the Hamiltonian with
the boundary contribution given above \cite{aaopen} suggested that,
it is not solvable by Bethe-ansatz techniques. The difficulty had to
do with the existence of the non-trivial boundary conditions which
led to non-factorizable interactions between the magnons and the
boundary.

It turns out that once the added boundary contributions are
included, the Hamiltonian (\ref{open22}) is solvable by the same
Bethe ansatz techniques that were applied in \cite{aaopen}! The
Bethe equations determine the magnon momenta $k_i$ by the
equations\beq \frac{\alpha (k_i)\beta (k_i)}{\alpha (-k_i)\beta
(-k_i)} = \prod_{j\neq i}\frac{S(-k_i,k_j)}{S(k_i,k_j)},\eeq where
\beq S(p,p') = \frac{\phi(p) - \phi(p') +i}{\phi(p)- \phi(p') -i},
\phi(p) = \frac{1}{2}\cot \left(\frac{p}{2}\right) \sqrt{1+ 8\lambda
^2\sin ^2\left(\frac{p^2}{2}\right)}.\eeq $\alpha, \beta $ remain
the same as they were at the one loop level, while the formula for
the bulk scattering matrix above is to be understood as accurate to
$O\lambda ^2$. The two loop dispersion relation being given by \beq
E(p) = 4\sin ^2\left(\frac{k}{2}\right) - 16\lambda \sin
^4\left(\frac{k}{2}\right).\eeq It is thus extremely encouraging to
see that integrability probably does continue to be present even in
the case of giant-graviton boundary conditions. Evidence for
classical open string integrability in the giant graviton
background, from the world-sheet point of view  has also recently
been presented in\cite{mann}. For a detailed analysis of various
other recently discovered aspects of open string dynamics and their
gauge theory duals we shall refer to\cite{maldaopen}. It would of
course be extremely interesting to comprehensively establish the
perturbative integrability of the open string sector of the gauge
theory from the point of symmetries and higher conserved charges
discussed before. The role of non-trivial boundaries/backgrounds in
our understanding of integrability of the gauge theory at higher
loops clearly are an exciting avenue
of exploration.\\

{\bf Acknowledgements:} We are extremely indebted to Diego Hofman
and Juan Maldacena for discussions and for sharing the manuscript of their recent
work\cite{maldaopen} before it was published. It is a great pleasure to thank
A.P.Polychronakos and S.G.Rajeev for their collaborations on
refs\cite{aaa} and\cite{aa1} respectively.

\end{document}